\documentclass{article}
\usepackage{waspaa21,amsmath,amsfonts,bm,graphicx,microtype,url,siunitx,times}
\usepackage{caption,subcaption,multirow,tabularx,booktabs,makecell}
\usepackage{mathtools} 
\usepackage[dvipsnames,svgnames,x11names]{xcolor}
\usepackage{color}

\input{def.set}

\DeclarePairedDelimiterX{\distx}[2]{(}{)}{%
  #1\;\delimsize\|\;#2%
}

\title{Zero-Shot Personalized Speech Enhancement\\through Speaker-Informed Model Selection}

\name{Aswin Sivaraman, Minje Kim\sthanks{This material is based upon work supported by the National Science Foundation under Grant No. 2046963.}}
\address{Indiana University, Department of Intelligent Systems Engineering, USA\\\small\tt{asivara@indiana.edu}, \tt{minje@indiana.edu}}

\begin{document}

\ninept
\maketitle

\begin{sloppy}

\begin{abstract}
This paper presents a novel zero-shot learning approach towards personalized speech enhancement through the use of a sparsely active ensemble model. Optimizing speech denoising systems towards a particular test-time speaker can improve performance and reduce run-time complexity. However, test-time model adaptation may be challenging if collecting data from the test-time speaker is not possible. To this end, we propose using an ensemble model wherein each specialist module denoises noisy utterances from a distinct partition of training set speakers. The gating module inexpensively estimates test-time speaker characteristics in the form of an embedding vector and selects the most appropriate specialist module for denoising the test signal. Grouping the training set speakers into non-overlapping semantically similar groups is non-trivial and ill-defined. To do this, we first train a Siamese network using noisy speech pairs to maximize or minimize the similarity of its output vectors depending on whether the utterances derive from the same speaker or not. Next, we perform k-means clustering on the latent space formed by the averaged embedding vectors per training set speaker. In this way, we designate speaker groups and train specialist modules optimized around partitions of the complete training set. Our experiments show that ensemble models made up of low-capacity specialists can outperform high-capacity generalist models with greater efficiency and improved adaptation towards unseen test-time speakers.
\end{abstract}


\section{Introduction}
\label{sec:intro}

Speech enhancement (SE) is a long-standing research area within signal processing
\cite{BollSF79ieeeassp, EphraimY1984spectralamplitude, Gannot98ieeesap}
which has experienced significant progress in the past decade due to the pervasiveness of machine learning models and deep neural networks (DNNs)
\cite{XuY2014ieeespl, HuangP2015ieeeacmaslp, WeningerF2015lvaica, WangDL2018ieeeacmaslp}. In this paper, we address the task of removing non-stationary background noise from a monophonic recording of a single speaker. Most studies in this area put forward models which perform generalizable speech enhancement---that is, the denoising models do not make assumptions about the test-time speaker or noisy environment. To address many kinds of speaker types and noise types, most DNN-based general-purpose models operate using millions of learnable parameters. However, given the ubiquity of resource-constrained devices (i.e., smartphones or smart speakers), we focus our interest on developing SE models which minimize run-time computational complexity without sacrificing denoising quality.

As it is known that models trained specifically for a disjoint sub-problem (i.e., \textit{specialists}) outperform the ones that aim at the universal speech enhancement task (i.e., \textit{generalists}) \cite{KolbakM2017ieeeacmaslp}, classifying which sub-problem the test signal belongs to can help improve the performance of the SE problem.  
A popular neural network design philosophy that handles those decomposed sub-problem individually is the mixture of local experts (MLoE) \cite{JacobsR1991nc}. 
An ensemble model consists of independent expert modules, each of which addresses a subset of all the training cases as a specialist. Hence, an auxiliary classifier module must be trained such that it estimates the ensemble weights for all the local experts based on their relevance towards a particular input example. The weights are then used to emphasize more relevant specialists' results. 

Recent studies applied the MLoE paradigm directly towards speech enhancement. In \cite{ZhangXL2016ieeeacmaslp} local experts are associated with different hyperparameter choices, such as different contextual window lengths, and then they are stacked up to form multicontext stacking.  MLoE is also extended to recurrent neural network models to learn from the temporal structure of the speech enhancement problem in \cite{ChazanSE2017MoERNN}.  While in these methods a local expert is not dedicated to solve a specific sub-problem, in \cite{SivaramanA2020interspeech} MLoE-based SE system shows substantial improvement by pre-defining two separate partitioning schemes: based on the quality of input signal, i.e., in terms of signal-to-noise ratio (SNR), and the gender of the speakers. Moreover, by introducing ``sparseness" to the ensemble weights, it performs test-time inference on only one most suitable specialist. Compared to an ordinary generalist that needs to be very large to achieve a certain level of speech denoising, each specialist can achieve the same quality enhancement results using much smaller architecture. Hence, the sparse ensemble of specialists is claimed as a model compression method in \cite{SivaramanA2020interspeech}. Zezario et. al also proposed a similar partitioning strategy based on the speech quality \cite{ZezarioRE2020zeroshotse}, but by employing a speech quality estimator in place of the traditional gating module. In \cite{ZezarioRE2020zeroshotse}, it is also proposed to learn the SNR-based partitions rather than pre-defining them. 

In this paper, we propose to use MLoE as a means for \textit{personalizing} the SE model. In doing so, instead of partitioning the dataset into pre-defined subsets, we take a more adaptive approach and learn the optimal speaker grouping strategy from the data.
With the speaker groups in place, the gating module must estimate characteristics of the test-time speaker (i.e., by computing an embedding), identify the most similar speaker group defined within the training set, then forward the input signal to the appropriate specialist network. 
This schema requires no training data from the test-time speakers but more optimally denoises the test-time noisy utterances by estimating the most suitable specialist. 
The idea of ``zero-shot'' speech enhancement through model selection has seen preliminary assessment, however, these works were limited to doing model selection based on the speaker-agnostic aspects of the signals, such as quality of the test-time signal, i.e. its signal-to-noise (SNR) level, \cite{ZezarioRE2019interspeech, ZezarioRE2020zeroshotse, SivaramanA2020interspeech}, types of the noise sources \cite{KimMJ2017icassp}, or the gender \cite{SivaramanA2020interspeech}.

Another important aspect of our work revolves around an open-ended question: how do we cluster English speakers into appropriate groups? This question is related to the task of learning speaker-characteristic embeddings, which has been extensively researched in the last decade, largely for improving speaker verification (SV) systems. Well-established embeddings include \textit{i-vectors}, which are calculated using a Gaussian mixture model \cite{DehakN2010ivector}, or \textit{x-vectors}  computed from a time-delay neural network \cite{SnyderD2017xvector}. Autoencoders' bottleneck layers are also a popular choice for learning fixed-length embeddings,
which have been used to compress Mel-frequency cepstral coefficients (MFCCs) into discriminative embeddings either through contrastive loss on pairs of inputs \cite{ChenK2011speakerspecific} or by estimating subsequent frames for a single input signal \cite{JatiA2017speaker2vec}. While these formulations are valid for learning speaker-identifying features, we propose to customize this embedding learning models further, so they effectively function as the gating module that is specialized for our SE problem rather than verification. To this end, we will first employ a Siamese network \cite{BromleyJ1994nips, ChiccoD2021siamese} to learn the discriminative speaker embeddings, and then turn it into a classifier as our gating module. Fine-tuning follows to further evolve the learned embedding space into something more suitable for the SE setup. 



\section{Proposed Methods}
\label{sec:methodology}

\subsection{Ensemble Models}

Given a large dataset containing many different speakers' various utterances $\mathbb{S}$, we postulate that there exists an optimal clustering within the data based around the speaker's identifying characteristics. Denoting $K$ to be the number of clusters, one can create $K$ separate SE models trained only to denoise utterances from each disjoint group of similar speakers. As previously shown \cite{SivaramanA2020interspeech, ZezarioRE2020zeroshotse}, a sparsely active ensemble model is capable of performing zero-shot adaptation because the gating module classifies the test-time \textit{noisy} input signals into one-of-$K$ groups.  


\begin{figure}[t]
    \centering
    \includegraphics[width=\linewidth]{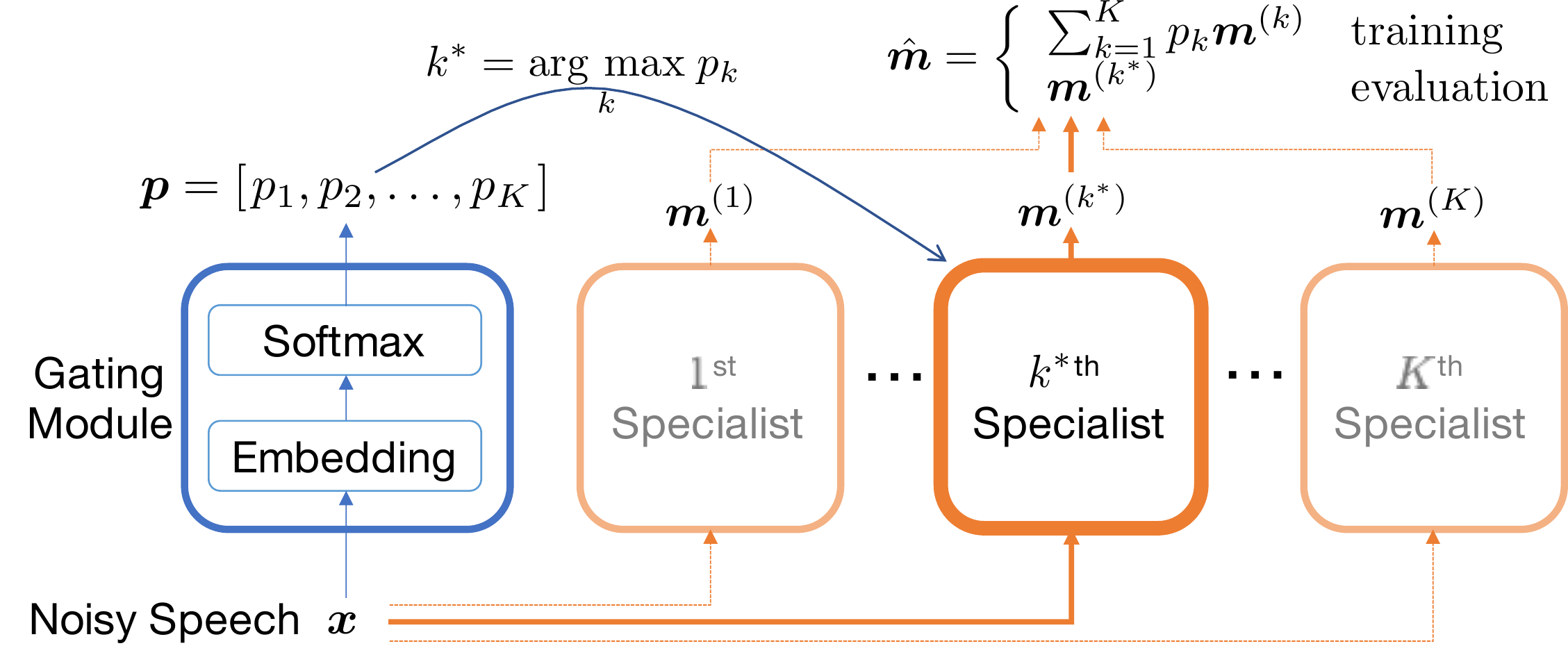}
    \caption{The proposed sparse ensemble of specialists model.}
    \label{fig:ensemble}
\end{figure}

We illustrate the architecture of a sparsely active ensemble model in Figure \ref{fig:ensemble}.
An ensemble model is composed of one gating module and $K$ specialist modules. The gating module processes a noisy speech input frame $\bm{x}$, estimating a speaker-embedding first and then classifying it as belonging to one-of-$K$ groups. The cluster probabilities vector $\bm{p}$ is used in two ways---during training, all of the specialist modules outputs their own ideal ratio mask (IRM) \cite{NarayananA2013icassp} estimates, $\bm{m}^{(1)}, \bm{m}^{(2)}, \ldots, \bm{m}^{(K)}$, which are then combined in a weighted sum using $\bm{p}$, i.e., $\hat{\bm{m}}=\sum_{k=1}^K p_k\bm{m}^{(k)}$. But during testing, only the output from the $k^{*}$-th specialist, corresponding to the largest probability, i.e., $k^{*}=\argmax_k p_k$, is chosen. This $\operatorname{argmax}$ operation selects a single specialist to use during evaluation, making the ensemble sparsely active.

In the context of personalized speech enhancement, increasing hyper-parameter $K$ can theoretically increase the level of specialization of each specialist as well as the ensemble network's capacity for personalization. However, there is a trade-off with having too many models; a large $K$ can make the gating module's classification task too challenging, and may lead to specialist modules becoming overfit on too small subset. In this paper, we investigate three choices of $K$: \si{2}, \si{5}, and \si{10}. Determining the optimal number of clusters is an extended research topic within unsupervised learning.

\subsection{Discriminative Speaker-Specific Embeddings}

The clustering of speakers is a significant matter when we build a successful sparse ensemble model for SE. Although in theory all the specialists and the gating module can be trained from scratch, training many modules simultaneously is prone to result in suboptimal performance. Hence, we first pre-train all the modules individually and then fine-tune them. The pre-training step, therefore, requires sub-grouping of speakers. 

To this end, we train a neural encoder that learns an an embedding function $f$ which can characterize a noisy speech utterance with a low-rank embedding vector.
In order to train $f$, we formulate a speaker verification (SV) upstream task. First, we sample utterances from a large training dataset containing many speakers, $\bm{s} \in \mathbb{S}$, and noise signals from a similarly large dataset of diverse noises, $\bm{n} \in \mathbb{N}$. Input mixtures $\bm{x}$ are made by artificially mixing clean speech utterances $\bm{s}$ with training noise signals $\bm{n}$; the amplitude of $\bm{n}$ is scaled to simulate various signal-to-noise ratios (SNRs).

We can then generate pairs of noisy speech utterances, $\bm{x}_{i}$ and $\bm{x}_{j}$. Once $f$ predicts the embeddings, i.e., $\bm{z}_i = f(\bm{x}_{i})$ and $\bm{z}_j = f(\bm{x}_{j})$, their inner product serves as a measure of similarity. A sigmoid function follows to interpret it as a probability $\hat{{y}}$. Our target is a binary value ${y}$, either \si{1} or \si{0} depending on whether the utterances derive from the same speaker or not. The embedding function $f$ is trained to minimize the binary cross entropy loss between $\hat y$ and $y$.

This contrastive learning approach for deriving discriminative embeddings follows the steps popularized by Siamese networks \cite{BromleyJ1994nips, ChiccoD2021siamese} in that the same embedding function $f$ is used for both input signals $\bm{x}_{i}$ and $\bm{x}_{j}$. The rationale behind this embedding model is that the discriminative nature of these embeddings can help the clustering process prepare a semantically more meaningful partitioning of speakers. 



\begin{figure*}[t]
     \centering
     \begin{subfigure}[b]{0.24\linewidth}
         \centering
         \includegraphics[width=\textwidth]{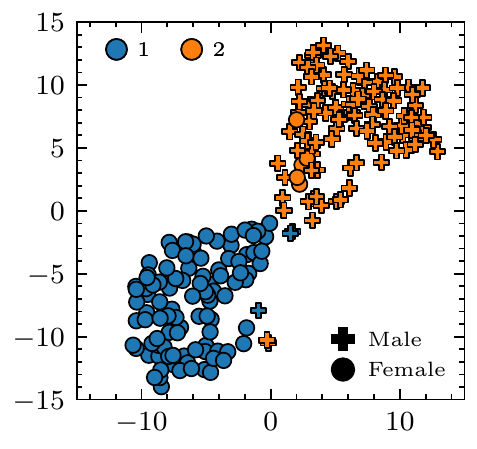}
         \caption{Speaker means obtained from SV with $K=2$ clustering.}
         \label{fig:tsne_k02}
     \end{subfigure}
     \hfill
     \begin{subfigure}[b]{0.24\linewidth}
         \centering
         \includegraphics[width=\textwidth]{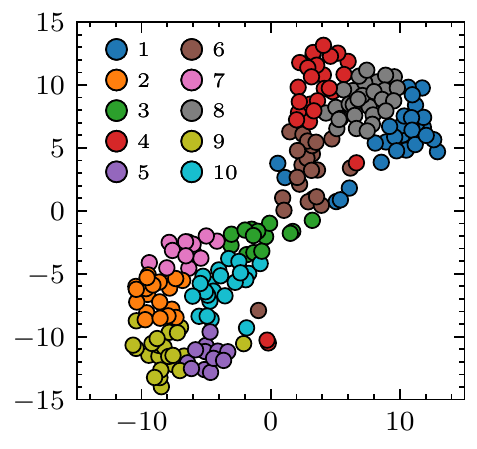}
         \caption{Speaker means obtained from SV with $K=10$ clustering.}
         \label{fig:tsne_k10}
     \end{subfigure}
     \hfill
     \begin{subfigure}[b]{0.24\linewidth}
         \centering
         \includegraphics[width=\textwidth]{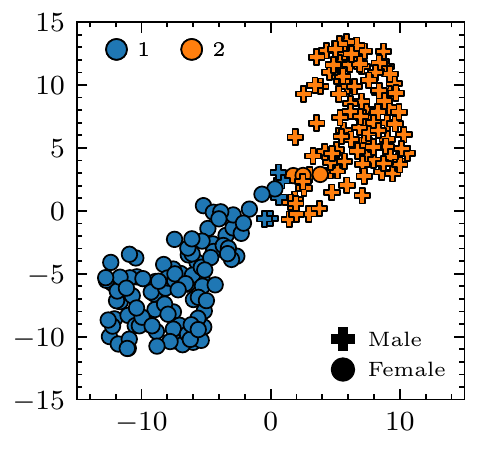}
         \caption{Speaker means derived uniquely by a fine-tuned $K=2$ ensemble.}
         \label{fig:tsne_k02ft}
     \end{subfigure}
     \hfill
     \begin{subfigure}[b]{0.24\linewidth}
         \centering
         \includegraphics[width=\textwidth]{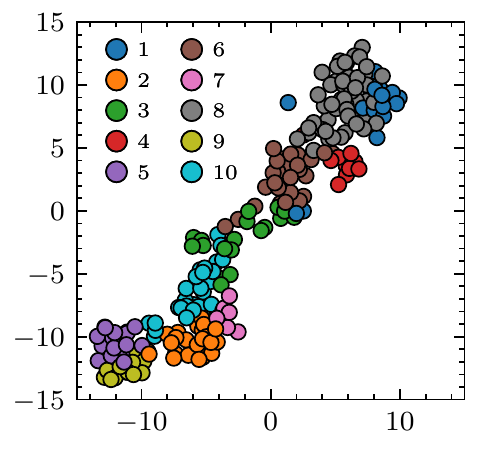}
         \caption{Speaker means derived uniquely by a fine-tuned $K=10$ ensemble.}
         \label{fig:tsne_k10ft}
     \end{subfigure}
        \caption{Subplots comparing various choices of $K$ for using k-means clustering on the speaker embeddings. The speaker verification (SV) pre-training task creates a latent space of speaker embeddings $\mathbb{Z}$, from which we can partition various groups, i.e. 2 in (a) and 10 in (b). After fine-tuning an ensemble model, the gating network's embedding function $f$ adjusts its parameters towards the speech enhancement (SE) objective. The latent space is modified uniquely based on the ensemble's configuration. In (a) and (b), the class labels derive from k-means clustering, but in (c) and (d) the class labels are estimated by the gating network's classifier function $g$.}
        \label{fig:tsne}
\end{figure*}

\subsection{Offline Speaker Clustering}\label{sec:kmeans}

Likewise, the gating module's classification task and pre-training of individual specialists rely on a reasonable clustering of speakers. 
Determining how the $K$ groups are formulated, and which of the training set speakers belong to each group, requires an offline clustering step. First, we transform every utterance from the training corpus into the learned latent space, i.e., $\bz\leftarrow f(s)$. Embedding vectors from the same speaker are averaged element-wise, which serves as the speaker-characteristic mean vector. Finally, we apply k-means clustering to these mean vectors, which learns $K$ speaker groups. 

Figure \ref{fig:tsne} illustrates the clustering results with different choices of $K$. Each of the \si{211} points represents one of the Librispeech training set speakers, with marker style indicating speaker gender. For plotting, the \si{32}-dimensional embeddings $\bz$ are reduced to \si{2} dimensions using t-SNE (with $\text{perplexity}=40$) \cite{VanDerMaatenL2008tsne}. These subplots show that the SV model succeeds in learning a speaker embedding which can be clustered into loosely meaningful groups, e.g., when $K=2$ the clusters implicitly form along the speaker gender division. These speaker groups are used to pre-train our gating modules and speaker-specific local experts. 

\subsection{Gating Module Pre-Training}\label{sec:gating}

The gating module must be able to classify the embedding vectors as belonging to one of the $K$ speaker clusters. This neural network is a dense layer followed by the softmax activation, which we denote by a parametric function $\bp=g(\bz; \mathcal{W}_g)$, where $\mathcal{W}_g$ is its parameters. The classifier function $g$ takes embeddings of noisy utterances $\bm{z}$ as inputs, and outputs a vector of cluster probabilities $\hat{\bm{p}}$. As each utterance belongs to a single cluster and the speaker IDs of the training set speakers are known, we can encode the k-means clustering labels into one-hot vector targets $\bm{p}$. These vectors are $K$-dimensional.

Note the discrepancy between the clustering done on embeddings of the clean speech utterances and the actual use-case of the model that takes noisy utterances. While the clustering results on clean data might be more reliable, eventually it is always possible that a noisy test utterance can be misclassified into a wrong speaker group, and then consequently assigned to a suboptimal specialist. Moreover, since the embeddings are optimized for the SV tasks, clustering on this representation may not be optimal for our SE problem. We revisit this issue in Sec. \ref{sec:finetune} and propose a fine-tuning solution.  

\subsection{Specialist Pre-Training}

The $K$ specialist modules are trained towards speech denoising in the following manner. The large dataset of training noises $\mathbb{N}$ is retained, but the large speech corpus $\mathbb{S}$ is partitioned into $K$ groups, based on the clustering results in Sec. \ref{sec:kmeans}. We denote the $k$-th partition as $\mathbb{S}^{(k)}$. The $k$-th specialist module can be described as a mapping function $h$ that updates its parameters $\mathcal{W}_h$ with each iteration such that the distance $\mathcal{E}$ between the denoised estimate signal $\hat{\bm{s}}$ and the target clean speech signal $\bm{s}$ is minimized. 
We use the negative scale-invariant signal-to-distortion ratio (SI-SDR) \cite{LeRouxJL2018sisdr} as the loss function.



\subsection{Ensemble Fine-Tuning}\label{sec:finetune}

The ensemble model can now be used na\"ively by assembling the pre-trained specialist modules and a pre-trained gating module. However, we note that the gating module may not classify all input signals with perfect accuracy.
Therefore, the ensemble model can benefit from fine-tuning (FT) in order to adjust denoising performance for misclassified inputs. This potential co-adaptation between gating and specialist modules can be found by adjusting the parameters of all the underlying functions (i.e., embedding function $f$, classifier function $g$, and denoising functions $h$ within each specialist). In the fine-tuning phase, the ensemble model estimates the final ratio mask $\hat{\bm{m}}$ by performing a normalized sum over the individual masks $\bm{m}^{(k)}$ using the softmax vector, $\hat\bp$, i.e., $\hat{\bm{m}} = \sum_{k=1}^{K} \hat{p}_{k} \bm{m}^{(k)}$. This ensures that the ratio mask calculation is differentiable and can be seen as a ``soft" gating mechanism.

During testing, the weighted sum is replaced by a hard-decision, i.e. $\hat{\bm{m}} = \bm{m}^{(k^\ast)}$ where $k^\ast = \operatorname*{argmax}_k p_k$. This switch in gating mechanism between training- and evaluation-time is the essence of the ensemble scheme's efficiency: only one out of all the specialists is active during inference, making the total used network parameters a fraction of the total learned. In order to reduce the discrepancy between the hard and soft gating mechanisms, a scaling parameter $\lambda = 10$ is applied to the logit values to make the softmax function harder. 

Figure \ref{fig:tsne}c and \ref{fig:tsne}d show the speaker embedding vectors after fine-tuning. Note that the comparison between the clustering on the SV embedding vectors and on their fine-tuned version is not to argue that fine-tuning can improve the clustering results. Instead, it is possible that fine-tuning with the speech enhancement objective could in fact deteriorate the discriminative qualities of the learned embedding vectors.  

\section{Experiment Setup}
\label{sec:experiment}

Mixtures are generated by combining randomly offset \SI{5}{\sec} segments of utterances and noises. With every mixture, the noise signal is randomly scaled such that the mixture SNR lies uniformly between \SI{-5} to \SI{10}{\decibel}. Utterances derive from the LibriSpeech corpus \cite{PanayotovV2015Librispeech} \textit{train-clean-100} folder, with \si{211} speakers designated in the training set, \si{20} in the validation set, and \si{20} in the test set. Noises are selected from the MUSAN corpus \cite{SnyderD2015MUSAN}, with \si{628} noises from the \textit{free-sound} folder used during training and validation, and \si{54} noises from the \textit{sound-bible} folder used during test. Both LibriSpeech and MUSAN corpora are resampled to \SI{8}{\kilo\hertz}. When training the speaker verification model, batches are made up of pairs of mixtures, with an equal chance of being from the same speaker or not.

All mixture signals are processed in the time-frequency domain through STFT using a frame size of \si{1024} samples with \SI{75}{\%} overlap. Throughout our experiment, every model performs speech denoising by taking a series of magnitude spectra as input and estimating IRM vectors $\bm{m}$. Masking is done element-wise onto the complex-valued spectrum which possesses the noisy phase of the mixture signal. 

Both the gating and specialist modules are composed of gated recurrent units (GRU) cells \cite{ChoK2014arxiv}. The embedding function $f$ is built with \si{2} hidden layers and \si{32} hidden units, with the output from last frame becoming a fixed-length utterance-characteristic embedding $\bm{z}$. The denoising functions $h$ are also built with 2 hidden layers but with a varied number of hidden units. The baseline general-purpose SE model is constructed in exactly the same manner as a specialist network, but is trained on the entire speech corpus $\mathbb{S}$ instead of a personalized subset $\mathbb{S}^{(k)}$. Throughout the experiment, we opt for a batch size of \si{128}, training all models using the Adam optimizer \cite{KingmaD2015adam} with learning rates of $10^{-3}$ for training and $10^{-4}$ for fine-tuning. 



\section{Results}
\label{sec:pagelimit}

\begin{figure}
    \centering
    \includegraphics[width=\linewidth]{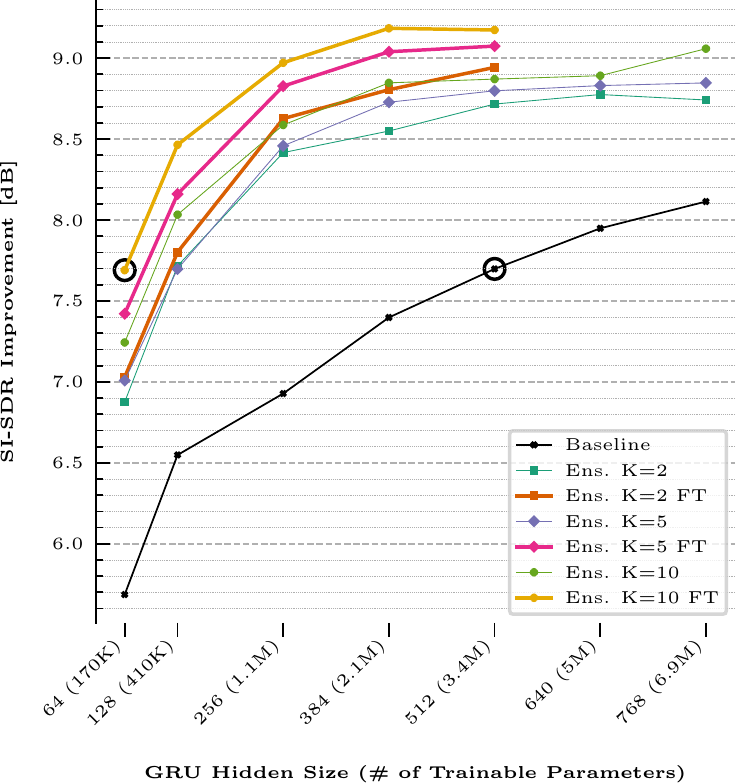}
    \caption{Comparison of speech enhancement performance between a baseline general-purpose model against different configurations of speaker-informed sparse ensemble models.}
    \label{fig:results}
\end{figure}


Figure \ref{fig:results} summarizes the findings of our experiments. The x-axis shows the varying hidden sizes for the GRU layers. In order to calculate the total number of parameters, the number in parenthesis must be multiplied by $K$, i.e., for the ensemble model with \si{5} specialists of hidden size \si{256}, the total number of parameters equals \SI{5.6}M. Because our ensemble models are sparsely active---that is, one specialist is active at a time---the number of parameters effective at run-time is only \SI{1.1}M, the amount listed on the x-axis. Longitudinally, the baseline models share the same number of hidden units with the specialist module, meaning the total number of model parameters for the baseline is always $K$ times smaller than the ensemble model in comparison. However their effective number of parameters is nearly equivalent. We note that ensemble models are not fine-tuned for hidden sizes $\geq \si{512}$ due to GPU memory constraints. Larger baseline models are trained and evaluated for comparison with the smaller ensemble models.

Firstly, we see that across all configurations, our ensemble models consistently yields a higher denoising performance when compared to a baseline generalist model whose size is similar to one of the specialists. The na\"ive ensemble models already show significant improvement (ranging from \SI{0.62} to \SI{1.65}{\decibel}), but different choices of $K$ do not make a big difference. We also observe that fine-tuning the ensemble models lift the performance even further (from \SI{1.24} to as much as \SI{2.04}{\decibel}. Furthermore, fine-tuning introduces a larger gap in improvement when $K$ is larger; intuitively, the more challenging classification task stands to benefit most from fine-tuning.

The proposed method also performs model compression without sacrificing the denoising performance. Overall, the smaller model architecture receives more performance improvement, such as the \SI{2.0}{\decibel} improvement in the case of \si{64} hidden units. The model compression benefits are made clear by comparing data points laterally. For example, as circled in Figure \ref{fig:results}, a generalist model requires at least \si{512} hidden units in order to match the performance of a fine-tuned ensemble model with \si{10} specialists each made up of GRUs with only \si{64} hidden units.
Including the cost of the gating module and all the other specialists that are not chosen, this is still a \si{48}{\%} reduction in terms of spatial complexity. Moreover, if we only count the gating module and one chosen specialist, it is a \si{94}{\%} reduction in effective parameters, greatly reducing arithmetic complexity during test-time. 

Lastly, as hypothesized, we see that increasing the number of clusters results can result in a more personalized speech enhancement so long as the ensemble model is fine-tuned. The average SI-SDR improvement achieved with the ensemble models increases along with $K$ from \si{2} to \si{5} to \si{10} through fine-tuning. 



\section{Conclusion}
\label{sec:pagestyle}

In this paper, we investigated model adaptation through selection (the ``mixture of local experts'' paradigm) as a means for zero-shot personalized speech enhancement. Our method is zero-shot as the ensemble models are never exposed to the test-time speaker or test-time environment; instead, the gating module analyzes the test signal to determine the most similar training cases and selects the most appropriate specialist for denoising. We obtain a speaker-informed gating module by pretraining it with a constrastive speaker verification task. The training cases are transformed to a learned latent space where they are clustered using k-means. By identifying more clusters and training more low-cost specialists, our ensemble models are able to adapt better to unseen speakers. Our findings reinforce the idea that sparse ensemble models can outperform general-purpose speech denoising models of a similar architecture, additionally reducing run-time computational complexity.

\newpage

\bibliographystyle{IEEEtran}
\bibliography{mjkim}

\end{sloppy}
\end{document}